\documentclass[sigconf]{acmart}

\usepackage{booktabs} 

\setcopyright{rightsretained}

\usepackage[nolist]{acronym}
\usepackage{enumitem}
\usepackage{xcolor}
\usepackage{svg}
\usepackage[normalem]{ulem}
\usepackage{multirow}
\graphicspath{{./images/}}
\DeclareGraphicsExtensions{.pdf,.png} 
\svgpath{{./images/}} 
\usepackage{colortbl}

\acmDOI{10.1145/3605098.3636132}

\acmISBN{979-8-4007-0243-3/24/04}

\acmConference[SAC'24]{The 39th ACM/SIGAPP Symposium on Applied Computing}{April 8--12, 2024}{Avila, Spain}
\acmYear{2024}
\copyrightyear{2024}

\acmArticle{4}
\acmPrice{15.00}

\acmBooktitle{The 39th ACM/SIGAPP Symposium on Applied Computing (SAC '24), April 8--12, 2024, Avila, Spain}

\begin{document}
\title{Hybrid Active Teaching Methodology for Learning Development}
\subtitle{A Self-assessment Case Study Report in Computer Engineering}
  
\renewcommand{\shorttitle}{Hybrid Active Teaching Methodology for Learning Development}

\author{Renan Lima Baima}
\orcid{0000-0002-6281-8153}
\affiliation{
  \institution{University of Luxembourg}
  \department{FINATRAX Research Group, SnT}
  \streetaddress{29, Avenue J.F Kennedy}
  \city{Kirchberg}
  \postcode{L-1855}
  \country{Luxembourg}
  }
\email{renan.limabaima@uni.lu}

\author{Tiago M. B. Caetano}
\orcid{0009-0001-0812-1291}
\affiliation{
  \institution{Instituto Superior Manuel Teixeira Gomes}
  \streetaddress{R. Dr. Estevão de Vasconcelos 33a}
  \city{Portimão}
  \country{Portugal}
}
\email{a22002128@mso365.ismat.pt}

\author{Ana Carolina Oliveira Lima}
\orcid{0000-0002-3829-7394}
\affiliation{
  \institution{Lusófona University}
  \department{COPELABS}
  \streetaddress{Campo Grande 376}
  \city{Lisboa}
  \country{Portugal}
}
\email{p6567@ismat.pt}

\author{Emilia Leal}
\orcid{0009-0001-6246-1473}
\affiliation{
  \institution{Universidad Nacional de Rosario}
  \department{Educaci\'{o}n}
  \streetaddress{Maipú 1065}
  \city{Santa Fe, Rosario}
  \country{Argentina}
}
\email{emilia.lima.leal@hotmail.com}

\author{Tiago Candeias}
\orcid{0000-0002-5254-6262}
\affiliation{
  \institution{Lusófona University}
  \department{COPELABS}
  \streetaddress{Campo Grande 376}
  \city{Lisboa}
  \country{Portugal}
}
\email{p4200@ismat.pt}

\author{S\'{i}lvia Pedro Rebou\c{c}as}
\orcid{0000-0002-8475-9748}
\affiliation{
  \institution{Lusófona University}
  \department{COPELABS}
  \streetaddress{Campo Grande 376}
  \city{Lisboa}
  \country{Portugal}
}
\email{p5974@ismat.pt}

\renewcommand{\shortauthors}{R. L. Baima et al.}

\begin{abstract}
 The primary objective is to emphasize the merits of active methodologies and cross-disciplinary curricula in Requirement Engineering. This direction promises a holistic and applied trajectory for Computer Engineering education, supported by the outcomes of our case study, where artifact-centric learning proved effective, with 73\% of students achieving the highest grade. Self-assessments further corroborated academic excellence, emphasizing students' engagement in skill enhancement and knowledge acquisition.  

\end{abstract}

%
%




\begin{acronym}
\acro{PBL}[PBL]{problem-based learning}
\acro{CU}{Curriculum Unit}
\acro{UML}{Unified Modeling Language}
\acro{RUP}{Rational Unified Process}
\acro{XP}{Extreme Programming}
\acro{CRM}{Customer Relationship Management}
\end{acronym}


\received{29 September 2023}
\received[revised]{17 November 2023}
\received[accepted]{05 January 2024}

\maketitle

\section{Introduction}

Our case study presents a distinctive hybrid active methodology, which combines established active methodologies highlighted in scientific publications~\cite{bernacki2021systematic}. These teaching techniques have significantly impacted engineering courses in higher education~\cite{mills2003engineering}. This study aims to explore a specific set of active methodologies from a theoretical perspective, identified using a diagnostic analysis of students, which considers their profiles, cultures, interests, and career aspirations, as well as their inclination towards working as part of a team or as individuals.

 Following the core purpose of analyzing and evaluating the influence of a combination of the mentioned innovative teaching methodologies on students' perceptions and experiences, the proposed research question (RQ):
\textbf{RQ: }\textit{How does a hybrid teaching methodology affect students' perceptions of their learning experiences compared to the effectiveness in skill development and knowledge acquisition?}
With this backdrop, to address this RQ, we contrast traditional teaching methods in Computer Engineering with emerging active learning techniques and interdisciplinary curriculum formats. Our proposal seeks to understand the alignment between self-assessment and instructor evaluation, offering insights into the role of critical thinking in education. 
This paper presents a case study on applying active methodologies in classrooms, focusing on a novel interdisciplinary curriculum format for Computer Engineering and Business Management programs. The curriculum integrates entrepreneurship activities, revamping requirements analysis, and bridging business management tools with user-centric approaches. 
\section{Methodology}
\label{sec:methodology}

  To better assess the student learning development and guide the teaching outcomes, the educational approach integrates the following key concepts:

\begin{enumerate}
    \item Active Learning Methodologies~\cite{coorey2016active}: Implementing active methodologies, such as~\ac{PBL}, made them more engaged, participative, and responsible for their learning.
    \item Integration of Technology~\cite{stearns2012integration}: Incorporating technology in the classroom helped students access information, collaborate on projects, and showcase their knowledge.
    \item Personalized Learning~\cite{bernacki2021systematic}: Provided opportunities for students to explore their interests and progress at their own pace, where each student received the attention they needed to reach their maximum potential via the mentorships.
    \item Support for Diversity and Inclusion~\cite{menezes2018diversity}: Thus, promoting an inclusive, diverse environment helped students from different backgrounds or abilities to present solutions more adjusted for the environment to which they are accustomed.
    \item Entrepreneurship and Innovation~\cite{corral2018design}: Encouraging students to cultivate an entrepreneurial mindset~\cite{kuratko2021unraveling} and innovative skills transformed them into problem solvers and agents of change in their communities, converting their actions into actionable outcomes.
    \item Industry Partnerships~\cite{faizi2021conceptual}: Collaborating with actual companies provided students with practical experiences to develop their system focused on the client's needs.
    \item Mentorship~\cite{holmes2018dimensions}: The programs connected students with experienced mentors, offering personalized guidance and valuable insights contributing to diversity, inclusion, and customized learning.
    \item Formative Assessment~\cite{roselli2006experiences}: The ongoing assessment enabled students to understand their progress and improve, opening them up to continuous learning advantages of understanding what they can immediately change.
    \item Local Community and Social Impact Projects~\cite{roshandel2011using}: Involvement in projects with social impact transformed students' perspectives on their role in society and inspired them to engage in the development from an external demand.
    \item Soft Skills Development~\cite{gonzalez2011teaching, gonzalez2020key}: By strengthening interpersonal skills such as leadership and teamwork, conducted considering their entrepreneurial profile, helped deliver pitches and other artifacts created throughout the course. They are not treated solely as students but as founders, recognizing the importance of developing entrepreneurial skills from an early stage.
\end{enumerate}

\section{Results}
\label{sec:results}

This case study began in the first semester of the academic year of 2022-23, with the "Software Engineering"~\ac{CU}, and extended into the second semester, encompassing the "Enterprise Systems Architecture and Business Management"~\ac{CU}. A cohort of 15 students participated in this study. Students were challenged to choose themes of technology-focused innovation for smart cities in the Algarve region. The curriculum was tailored to this theme, and students' progress was monitored in teams.

A consistent pattern of active student participation was observed, exemplified by a class session with 26 interactions involving 13 students. Additionally, all participants displayed an eagerness to complete their tasks. Further analysis and description of their reporting and assessment are provided here.

A survey was administered at the end of the~\ac{CU} to gauge students' perceptions of the teaching methodologies and content. The survey utilized the Likert scale~\citet{likert1932technique}, which spans from 1 to 5, and open-ended questions. Out of the cohort, we received 15 responses.

\begin{figure}
    \centering
    \includegraphics[clip, trim=1.5cm 2cm 1.5cm 2cm, width =0.47\textwidth]{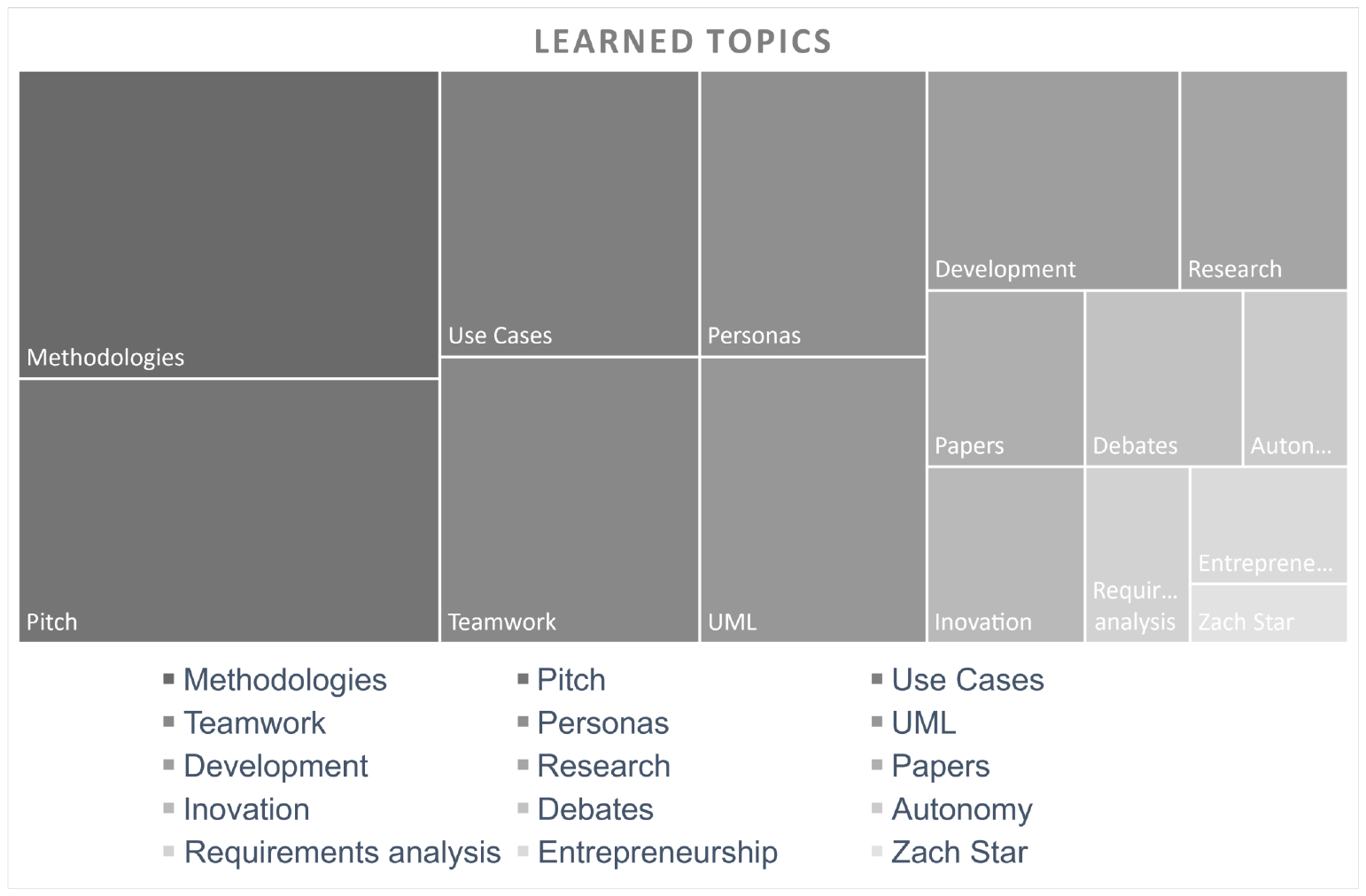}
    \caption{Learned Topics}
    \label{fig:LearnedTopics}
\end{figure}

Fig.~\ref{fig:LearnedTopics} showcases the topics students felt most comfortable with by the end of the~\ac{CU}. First, we can address the most learned topics, such as Methodologies by 17~\%, Pitch by 15~\%, Use Cases by 10~\%, and Teamwork by 10~\%, followed by the lesser learned reported topics as Zach's Star by 1~\%, Entrepreneurship by 2~\%, Requirements analysis by 2~\%, and autonomy by 2~\%. While it doesn't confirm mastery, it provides insights into students' perceived gains. It's important to clarify that this graph comprises open-ended answers so students can transmit all the subjects they deem to have learned.

\begin{figure}
    \includegraphics[clip, trim= 2.8cm 17.5cm 2.8cm 2cm, width =0.47\textwidth]{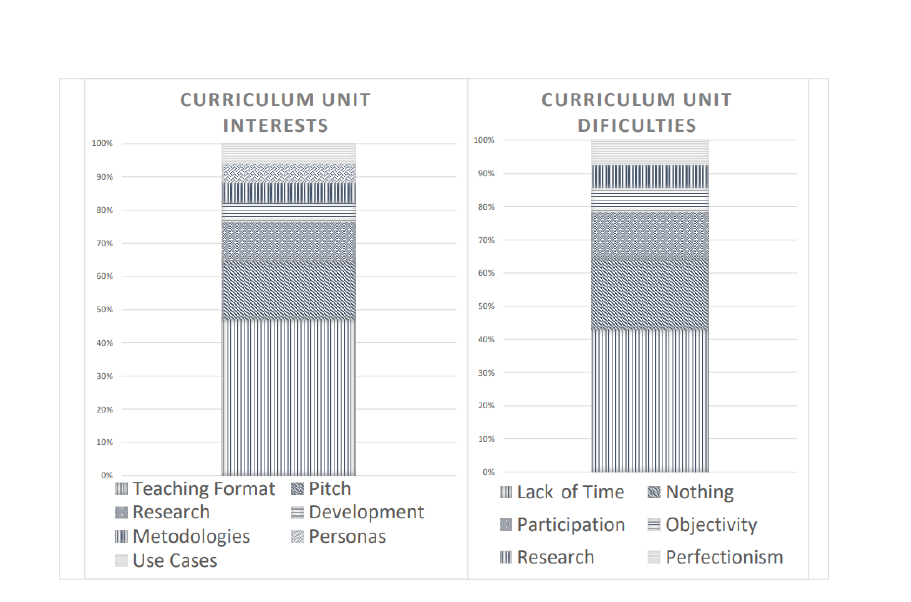}
    \caption{Student's Interests \& Difficulties}
    \label{fig:Interests&Dificulties}
\end{figure}

Fig.~\ref{fig:Interests&Dificulties} presents a dual perspective by highlighting topics that intrigued students and identifying challenges they encountered. This data can aid in refining the curriculum to better cater to students' interests and address their difficulties. Therefore, it provides a comprehensive view of the student experience, ultimately enhancing their learning and effectively addressing potential challenges. It is also important to note that the percentages in the graph represent the number of answers the students gave, as this was an area where they could provide more than one answer.

It's important to note that this distribution of interests is based on individual student contributions, where each had the freedom to mention the topics that interested them the most. The~\ac{CU} offered diverse topics that capture students' interests, including teaching format, pitch, research, development, methodologies, personas, and use cases. These interests reflect the students' desire for engaging and compelling learning experiences, effective communication, research exploration, practical application, structured problem-solving, and understanding user profiles. This variety of topics reflects the diversity of students' interests in the~\ac{CU}, contributing to a more enriching and comprehensive educational experience.

\section{Discussion}
\label{sec:discussion}

The integration of entrepreneurship into the Software Engineering curriculum has proven to be a practical approach, despite using entrepreneurship methodologies it did not necessarily lead students to focus on learning entrepreneurship but also more technical subjects, as they reported.

This supplementary material, blending entrepreneurship with Software Engineering, delivered a more holistic education and equipped students to confront the workforce's demands. As they honed their entrepreneurial and applied their technical expertise in practical scenarios, students became more adept at innovation, crafting software solutions, and devising business systems efficiently and according to market requirements.

The positive outcomes observed in our study align with the broader academic consensus on the benefits of an \ac{PBL} approach. Students' enthusiasm and positive reception of this approach underscore the importance of real-world problem-solving and the value of facilitation in the learning process. This is consistent with findings from other studies~\cite{fioravanti2018integrating, dos2016pbl}, emphasizing skills and competency-focused learning.

\section{Conclusions}
\label{sec:conclusions}

In this research, the use of active methodological approaches directly impacted both the software deliverables and high motivational interest for the elaboration of the students' ideation, led by a strong interest in entrepreneurship. Furthermore, these students achieved high-level skills and competencies, especially in communication during pitch presentations, delivering improved artefacts~\cite{bastias2021evaluation} and diagrams focusing on the end-user. %

In stark contrast to traditional Software Engineering teaching methods~\cite{glass2002research}, which rely on conventional techniques such as defining functional and non-functional requirements, using languages like UML, and choosing software development methodologies~\cite{kuratko2021unraveling, sommerville2020engineering}, the hybrid methodology approach greatly emphasizes entrepreneurship and innovation in social, regional, and community development, as per the students' subjective satisfaction responses. This approach highlighted the importance of adapting to the industry's ever-changing demands and providing continuous support to students to help them succeed.

\bibliographystyle{ACM-Reference-Format}
\bibliography{bibliography} 

\end{document}